%% file: safecert08.tex
\documentclass{entcs} \usepackage{prentcsmacro}
\usepackage{graphicx}
\def\lastname{Braun, Philipps, Sch\"atz, Wagner}
\begin{document}
\begin{frontmatter}
  \title{Model-Based Safety-Cases for Software-Intensive Systems} 
  \author{Peter
    Braun\thanksref{peter}}
  \author{Jan Philipps\thanksref{jan}}
  \address{Validas AG\\
    M\"unchen, Germany} 
  \author{Bernhard Sch\"atz\thanksref{barnie}} 
  \author{Stefan Wagner\thanksref{stefan}}
  \address{Institut f\"ur Informatik\\Technische Universit\"at M\"unchen\\
    Garching b.~M\"unchen, Germany}
    \thanks[peter]{Email:
    \href{mailto:peter.braun@validas.de} 
         {\texttt{\normalshape peter.braun@validas.de}}}
    \thanks[jan]{Email:
    \href{mailto:jan.philipps@validas.de} {\texttt{\normalshape 
    jan.philipps@validas.de}}}
    \thanks[barnie]{Email:
    \href{mailto:schaetz@in.tum.de} {\texttt{\normalshape
        schaetz@in.tum.de}}}
    \thanks[stefan]{Email:
    \href{mailto:wagnerst@in.tum.de} {\texttt{\normalshape
        wagnerst@in.tum.de}}}

\begin{abstract} 
Safety cases become increasingly important for software certification.
Models play a crucial role in building and combining information for the
safety case. This position paper sketches an ideal model-based safety case
with defect hypotheses and failure characterisations. From this, open
research issues are derived.
\end{abstract}

\begin{keyword}
Safety case, model-based, structured argument, defect hypothesis, failure characterisation
\end{keyword}

DOI:10.1016/j.entcs.2009.09.007

\end{frontmatter}

%%%%%%%%%%%%%%%%%%%%%%%

\input{intro}

\input{safetycase}
\input{models}

\input{issues}

\input{conclusion}

%%%%%%%%%%%%%%%%%%%%%%%

\bibliographystyle{plain}
\bibliography{safecert08}

\end{document}

%% file: intro.tex
\section{Introduction}\label{sec:intro}

The proliferation of software-intensive technical systems has resulted
in a growing need for methods to demonstrate their safety and
reliability. The goal of such methods is to develop a \emph{safety
  case} for a system -- a line of argument that establishes
%(quantitative or
%qualitative) 
safety and reliability properties from known properties of the
components of the system. 

% To that end, a model of the system is
% constructed--either partly implicitly as, e.g., in FTA, or explicitly
% as, e.g., in formal verification.
%
%The best-known methods are Failure Mode and Effects Analysis (FMEA)
%and Fault Tree Analysis (FTA), which build a line of argument starting
%from assumed component failures towards their effects on the systems
%or from undesired system failures backwards towards possible causes of
%these failures, respectively. The forward and backward analysis
%directions can also be combined, as in the HAZOP method. While these
%approaches are most effective when only hardware-related failures are
%considered, they have also been applied to software code and models.
Various approaches exist: Leveson et al.~\cite{leveson91} describe
an FTA-like approach to examine Ada programs, while Giese et al.\ 
\cite{giese04} show how HAZOP-like safety analyses can be based on
component and deployment diagrams of the UML. Within the ISAAC project
\cite{akerlund:isaac}, models of functional, geometrical and human
aspects are integrated for safety analyses. Pumfrey \cite{pumfrey99}
gathers a list of nine factors for success of safety analysis methods
and goes on to develop two methods for dealing with mixed hw/sw
systems.

\begin{figure}
  \begin{center}
    \includegraphics[scale=.7]{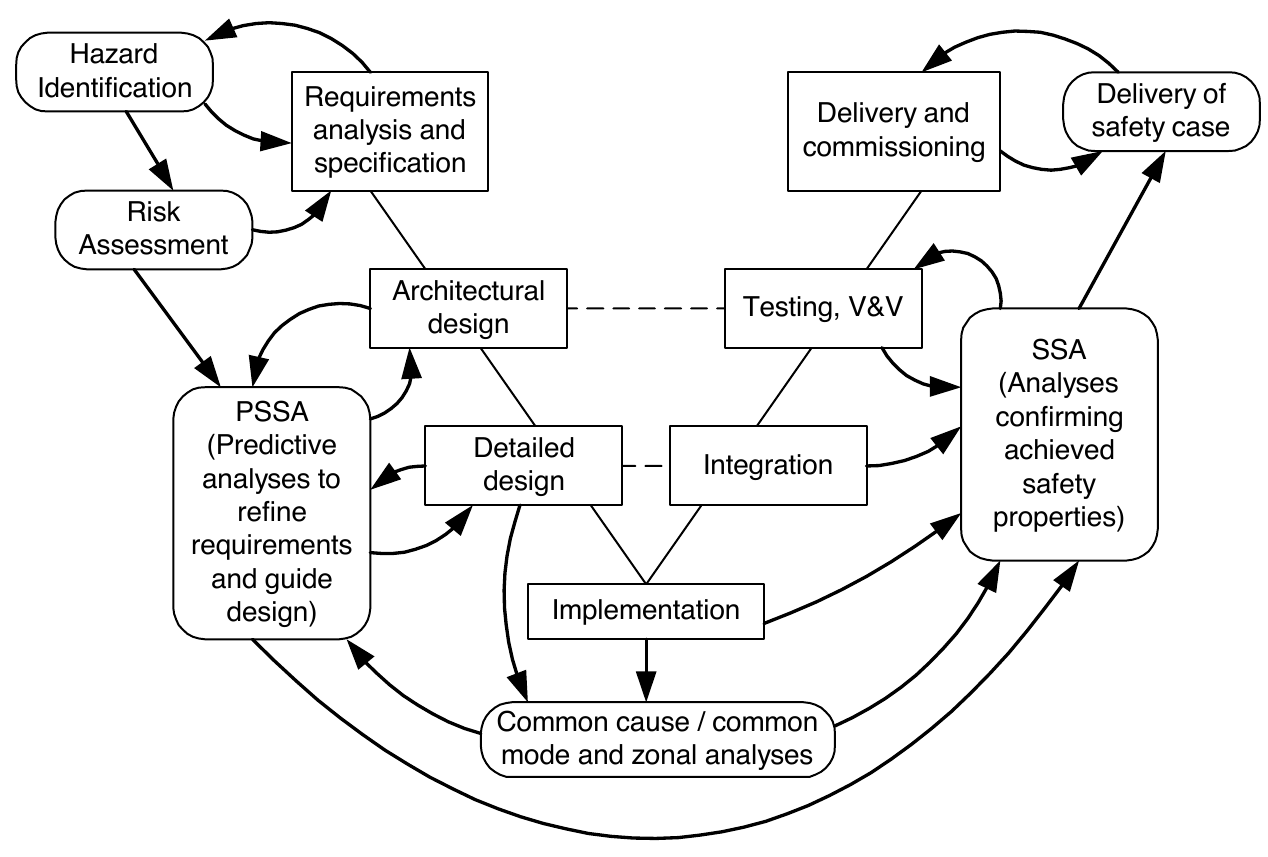}
    \caption{Safety activities in a development process (Source: \cite[p. 29]{pumfrey99})
      \label{fig:activities}}
  \end{center}
\end{figure}

% While obviously, the use of models plays a central role in the
% construction of a safety case, a systematic approach to the definition
% of those models is still uncommon. Especially in formal approaches
% often the more critical part of constructing the underlying model, and
% justifying its appropriateness for the safety case is often left
% unconsidered. 

%The
%classical methods in current use are essentially structured system
%reviews that make use of \emph{ad hoc}-descriptions of the system and
%the creativity of a small number of system engineers. It is not clear
%how they scale to modern software-intensive systems with their
%multitude of failure modes and still remain both trustworthy and
%cost-effective. 

In all these approaches the use of models plays a central role in the
construction of a safety case. While earlier approaches are based on
structured reviews of models, recently formal verification techniques
have been applied for model analysis
\cite{bozzano:esacs,giese04,joshi05,akerlund:isaac}. However, a
systematic approach to the definition of those models is still
uncommon. It is also an open issue how to justify the appropriateness
of the underlying models for the safety case: Is it possible to derive
all relevant hazards, system failures and component faults, and is it
possible to reason about the causal chains that link them?  In other
words, we believe that the major open issue is how to reason about the
\emph{choice} of models, and not so much how to reason about the
\emph{properties} of the models.

In addition to this principal issue of the appropriateness of the
models, we believe that there are a number of core success factors for
building model-based safety cases:
\begin{itemize}
\item \emph{Seamless integration into development processes.} It is not
  sufficient to merely perform a single safety analysis for
  certification of the final system -- analyses with different focus
  play their role throughout system development, in order to clarify
  requirements, designs, and in general to improve both product and
  process (see Fig.~\ref{fig:activities}).
\item \emph{Consideration of system, platform and environment.} It is
  not sufficient to examine models for the functional behaviour (even
  if they are augmented with fault models, as in \cite{joshi05}) of a
  system by verification or tests. Since hazards manifest themselves
  in the system's environment, the environment must also be modelled
  and included in the analysis. Since faults often are caused by the
  underlying computing platform, the platform must also be included;
  possibly, abstract user models may be needed to reason about
  operator errors and ways to avoid them or to deal with them. Note
  that in the development process these different models may well be
  constructed and analysed at different times: For instance, a
  preliminary hazard analysis (corresponding to the top left activity
  in Figure~\ref{fig:activities}) may need only rather abstract
  environment and system models. 
% Note that the interfaces between these models must be defined in
% order to reason about the propagation of fault effects. Giese et.
% al. \cite{giese04} propagate faults through the ports and connectors
% of deployment diagrams; it is not clear how fault propagation
% through heterogeneous causal chains (e.g., from software via the
% platform to a different software component) can be handled.
\item \emph{Heterogeneous reasoning.} Finally, in order to cope with
  the realities of systems development practice, where different
  components -- some newly developed, some legacy, some
  off-the-shelf -- of different sources are combined, a compositional
  approach to the development of safety cases with a mixture of
  quantitative and qualitative reasoning is needed. While some
  properties may be demonstrated through rigorous verification or
  testing, others may be based on statistical reasoning (as in some
  applications of FTA) and empirical data.
\end{itemize}

In this paper we outline a research agenda for model-based safety
cases that tries to give answers for some of these issues. In
Section~\ref{sec:case}, we give a short overview over the
argumentation behind a safety case and in Section~\ref{sec:models} we
look at the requirements on models and modelling languages that can
support the building of safety cases. Section~\ref{sec:issues} lists
some of the open research issues, and Section~\ref{sec:conclusions}
concludes.

%% file: safetycase.tex
\section{Safety Case}\label{sec:case}

\begin{figure}
\begin{center}
\includegraphics[width=.8\textwidth]{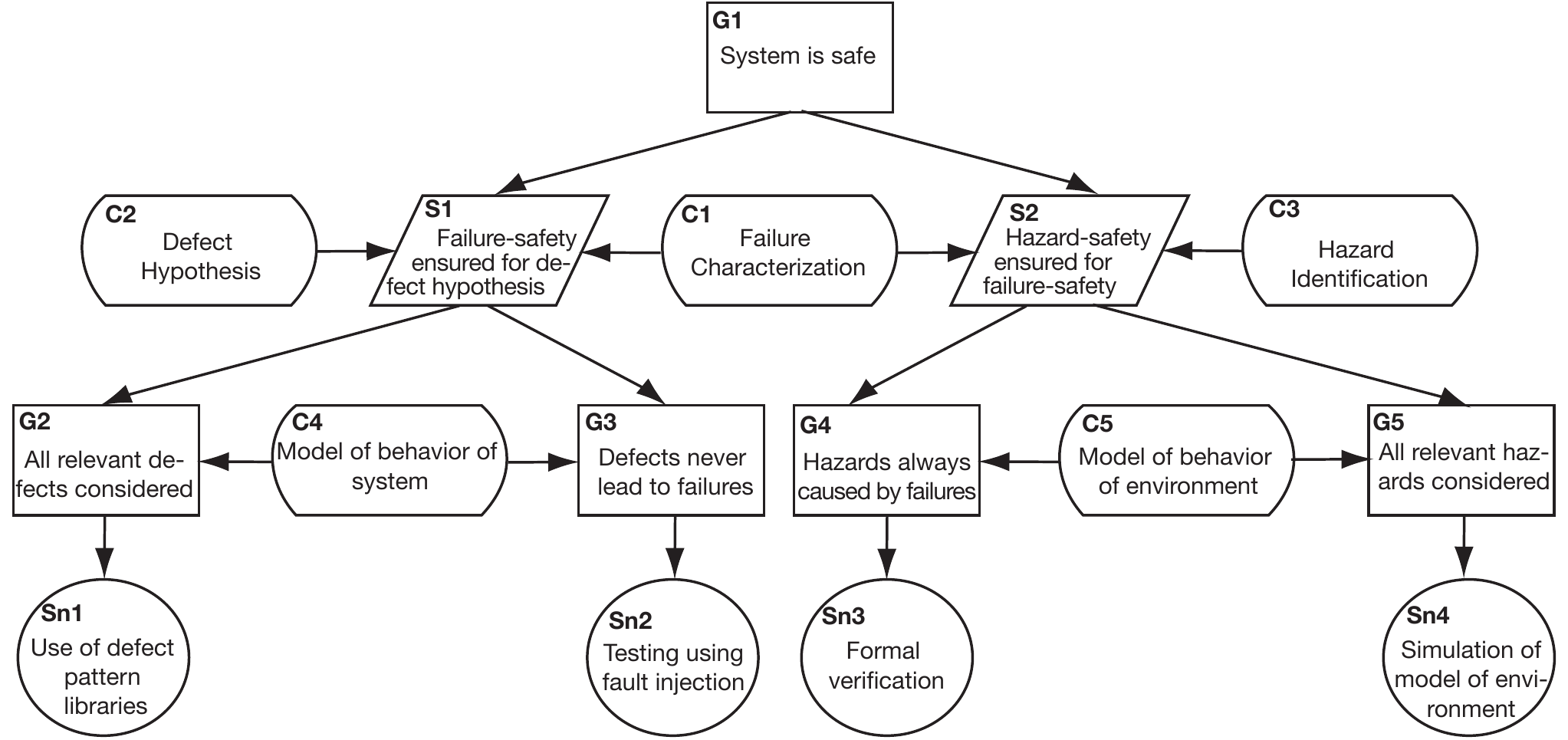}
\caption{A simplified example of a safety argument using 
         GSN and context information from models\label{fig:gsn-example}}
\end{center}
\end{figure}

In general, a safety case is a structured line of arguments that shows
that the system under consideration is safe.  One of the difficulties
is that a large variety of information needs to be combined to form
this argument. A single safety assurance method is never able to show
the complex issue of safety completely.  Hence, formal verification,
statistical testing, process conformance and other information must be
integrated for a convincing argument.

A description technique that has proven to be useful for constructing
safety arguments is the \emph{Goal Structuring Notation} (GSN)
\cite{kelly04}. It reduces some problems, such as ambiguity, of purely
textual descriptions.  An example safety argument using GSN is shown
in Fig.~\ref{fig:gsn-example}.  In this example the overall goal
\textbf{G1} is that the system is safe. This is intended to be
achieved by two strategies: \textbf{S1} is to ensure there are no
failures in terms of deviations from the intended safety-critial functionality, and 
\textbf{S2} arguments by showing that there are also no hazards in the absence of failures.
Hence, defect hypotheses as well as possible
hazards must be identified. This is depicted by the two constraints
\textbf{C2} and \textbf{C3}. From the strategies, four refined goals
are derived. \textbf{G2} expresses that all relevant defects need to
be considered,
which is shown by using defect pattern libraries in solution
\textbf{Sn1}. \textbf{G3}, \textbf{G4} and \textbf{G5} describe further
goals for failures and hazards, which are met by the exemplary solutions
\textbf{Sn2}, \textbf{Sn3} and \textbf{Sn4}. Important for the following
is also that there are several context informations like \textbf{C1}, the
characterisations of failures, and models of the behaviour of the system
(\textbf{C4}) and the environment (\textbf{C5}).

In summary, the example shows that a variety of information needs to
be considered when constructing a safety case. From a methodological
point of view, safety requirements and hazards need to be identified.
However, this is not sufficient. It is also necessary to build models
that form the context for more detailed arguments, for example about
specific components.  Moreover, the models must be able to handle
deterministic as well as probabilistic information and need not only
to include the system itself but also its environment.

%% file: models.tex
%SourceDoc ./safecert08.tex

\section{Model-Based Safety Assurance}\label{sec:models}

\emph{Model-based development} is becoming a common-place approach to
embedded (control) software construction; models, which describe the nominal functionality, are found in form of plant models, e.g., for in-the-loop testing, as well as controller models, e.g., for production-code generation. 
In some areas, the generation 
of production code from models is already state-of-the-art. %By means of
%\emph{compositional} modelling paradigms---i.e., modelling techniques 
%that allow to deduce the property of a composed functionality from the 
%properties of those sub-functionalities it is composed of---efficiency of e
%mbedded software development has been improved. 
%As already described in Section \ref{sec:intro} safety cases often make use of
%\emph{ad hoc}-descriptions. Therefore, the question arises how the
%model-based development approach can be extended appropriately to
%support a cost-effective and trustworthy hierarchical safety case
%construction as described in Section \ref{sec:case}. In the following
%we propose some relevant requirements for models to support this
%idea.

However, without specifically addressing the issue of safety-cases, models of
the \emph{nominal functionality} describing the system under
development are not sufficient for the analysis of fail-safe behaviour. To
apply the models used in a model-based development approach to the
construction of a safety-case,
\begin{itemize}
\item a model of the system must be derived to \emph{describe the effective functionality, including nominal as well as defect-affected behaviour}.
\item a model of the environment must be constructed to \emph{explicitly model the assumptions about the behaviour of the environment}.
\item explicit hypotheses of defects -- expressed solely in terms of the system -- must be provided to \emph{avoid mistaking fail-safe behaviour of the model for fail-safe behaviour of the system}.
\item explicit characterisation of failures -- expressed solely in terms of the interface between system and environment -- must be provided to \emph{allow identifying deviations from the intended behaviour in the interaction between the system and its environment}.
\item explicit identification of hazardous situations -- expressed solely in terms of the environment -- must be provided to \emph{allow describing hazards in terms of the controlled environment rather than the controlling system}.
\end{itemize}
Therefore, when extending the construction of the nominal
functionality of a system to the construction of a safety-case, the issues of
\emph{defect hypotheses}, \emph{failure characterisations}, as well as a \emph{hazard identifications}
must be addressed in the safety case, making implicit assumptions explicit. % either in form of constraints.% (e.g., as defect hypothesis like \textbf{C3} in Section \ref{sec:case}) or in form of safety goals like \textbf{G2}).

A failure characterisation describes unintended functionality of the
system, which may potentially lead to hazardous situation. In the
safety-case it is used to link the line of argument between the
environment model and the system model. 

A defect hypothesis describes how the nominal functionality of a system is altered to reflect the possible occurrences of faults. As defects may be introduced in different parts of a system, different
defect hypotheses are needed, e.g, to reflect defects caused by a deviation from the intended functionality of
  the system platform (i.e., hardware defects), or faults caused by a deviation from the intended control
  functionality (i.e., design defects). For practical application, defect hypotheses are often described in
form of fault patterns, e.g., in form of intermittent occurrences of value changes to reflect influences of alpha radiation. 

A hazard identification describes states of the controlled environment to be avoided by the nominal functionality of the system. However, as noted by \cite{Jackson:1996} and \cite{ParnasMadey:1995}, especially in the context of embedded or software-intensive systems, the functionality of the system is often only adequately expressed
over the part of the environment, which is only indirectly controlled or monitored by the system under development. Therefore, to describe the achievement and failure of functionality, a hazard identification independent of the model of the system is needed.

Separating these three properties as well as using separate models of system and environment reduces the complexity of the overall argument and minimises the risk of constructing inadequate lines of arguments.

As shown in Figure \ref{fig:gsn-example}, based on these five models, now a formalised and standardized line of argument can be constructed to  \emph{show the absence of hazardous situations based on the defect hypotheses}. This line of argument is constructed in six steps:
\begin{enumerate}
\item The defect hypothesis is validated to ensure that all relevant defects are included.
\item The hazard identification is validated to ensure that all relevant hazards are included.
\item The model of the system is verified with respect to conformance of the defect hypothesis, i.e., the \emph{system does not constrain the possibility of such defects}.\label{defect-hypothesis}
\item The model of the environment is verified with respect to conformance of the hazard identification, i.e. the \emph{environment does not constrain the possibility of such hazards}.\label{hazard-identification}
\item The model of the system is verified with respect to the avoidance of failure situations under the given defect hypothesis.\label{no-failures}
\item The model of the environment is verified with respect to the avoidance of hazardous situations under the given absence of failures.\label{no-hazards}
\end{enumerate}
For each of these steps, different solution techniques can be applied. 
 E.g., applying defect patterns for the identified defects can be used in step \ref{defect-hypothesis}, while a simulation of the environment model can be used in step \ref{hazard-identification}. 
Furthermore, steps \ref{no-failures} and \ref{no-hazards} themselves are complex verification goals, requiring appropriate sub-goals and sub-solutions.

%% file: issues.tex
\section{Open Research Issues}\label{sec:issues}

% We are aware that the described model-based safety case is far from
% being fully developed. Therefore, we summarise the most important
% open issues that need to be addressed.

Based on the ideas of the previous section, we briefly state the most
important open issues that need to be addressed.

\textbf{Safety Case.} To start with, how can safety requirements and
possible hazards be effectively elicited? When is the safety case complete?
A strong structure needs to be provided for the safety engineering in
the form of safety patterns and reusing existing parts of safety cases.
Moreover, the approach has to be mapped to existing (an potentially new)
standards to allow certification.

\textbf{Models.} In general, the question is how can be assured that
the built models are suitable for safety analysis. The quality of
these models is decisive for the whole safety case. There is a
plethora of questions: On what level of granularity are the models
built? What are suitable interfaces between components so that complex
causal chains can be analysed? What are proper defect hypotheses and
failure characterisations? Too much detail is not manageable, not
enough detail leads to omissions.

\textbf{Supporting Methods.} Finally, various and diverse methods need
to be used for arguing in the safety case. Hence, quantitative and
qualitative as well as deterministic and probabilistic methods need to
be used and integrated in the argument. When is formal verification
needed and feasible, when are other arguments necessary and
sufficient?

%% file: conclusion.tex
\section{Conclusions}\label{sec:conclusions}

Safety cases become increasingly important for software-intensive systems.
The current state-of-practice, FTA and FMEA, are not sufficient for the
enormously complex and interconnected modern systems. Hence, we need
suitable models, not only of the system but also of its environment,
especially its users. These models are needed as basis for formal
verification. Moreover, they feed into the complete and structured argument in
a safety case. We described how model-based safety analysis and safety case
development should ideally look like and derived a set of open research
issues that need to be addressed. We are currently in the process of
discussing with industrial partners first steps to tackle these issues.

%% file: safecert08.bbl
\begin{thebibliography}{1}

\bibitem{bozzano:esacs}
M.~Bozzano et. al.
\newblock {ESACS}: an integrated methodology for design and safety analysis of
  complex systems.
\newblock In {\em Proc. ESREL 2003}, pages 237--245, 2003.

\bibitem{akerlund:isaac}
O.~Akerlund et. al.
\newblock {ISAAC}, a framework for integrated safety analysis of functional,
  geometrical and human aspects.
\newblock In {\em Proc. ERTS 2006}, 2006.

\bibitem{giese04}
Holger Giese, Matthias Tichy, and Daniela Schilling.
\newblock {C}ompositional {H}azard {A}nalysis of {U}{M}{L} {C}omponents and
  {D}eployment {M}odels.
\newblock In {\em Proc.~23rd International Conference on Computer Safety,
  Reliability and Security (SAFECOMP)}, volume 3219 of {\em LNCS}. Springer
  Verlag, 2004.

\bibitem{Jackson:1996}
Michael Jackson.
\newblock {\em {Software Requirements and Specifications}}.
\newblock Addison-Wesley and ACM Press, 1996.

\bibitem{joshi05}
Anjali Joshi, Steven~P. Miller, Michael Whalen, and Mats~P.E. Heimdahl.
\newblock A proposal for model-based safety analysis.
\newblock In {\em Proc. 24th Digital Avionics Systems Conference}, Oct 2005.

\bibitem{kelly04}
Tim Kelly and Rob Weaver.
\newblock The goal structuring notation -- a safety argument notation.
\newblock In {\em Proc.~DSN 2004 Workshop on Assurance Cases}, 2004.

\bibitem{leveson91}
Nancy~G. Leveson, Stephen~S. Cha, and Timothy~J. Shimeall.
\newblock Safety verification of ada programs using software fault trees.
\newblock {\em IEEE Softw.}, 8(4):48--59, 1991.

\bibitem{ParnasMadey:1995}
D.~Parnas and J.~Madey.
\newblock {Functional Documents for Computer Systems}.
\newblock {\em Science of Computer Programming}, 1(25):41--61, October 1995.

\bibitem{pumfrey99}
David~John Pumfrey.
\newblock {\em The Principled Design of Computer System Safety Analyses}.
\newblock PhD thesis, Department of Computer Science, University of York, 1999.

\end{thebibliography}
